\documentclass[amsmath,amssymb,12pt]{article}
\usepackage[all]{xy}
\usepackage[centertags]{amsmath}
\usepackage{latexsym}
\usepackage{amsfonts}
\usepackage{amssymb}
\usepackage{amsthm}
\usepackage{graphicx}
\usepackage{dcolumn}% Align table columns on decimal point
\usepackage{bm}% bold math

\begin{document}

%\preprint{APS/123-QED}

\title{Acoustic cloaking and mirages with flying carpets}

%\author{Andr\'e \surname{Diatta}}
%\affiliation{University of Liverpool. Department of Mathematical
%Sciences, Peach Street, Liverpool L69 3BX, UK}
%\email{adiatta@liverpool.ac.uk}
%\author{Guillaume \surname{Dupont}}
%\affiliation{Institut Fresnel-CNRS (UMR 6133), University of
%Aix-Marseille, 13397 Marseille cedex 20, France}
%\author{S\'ebastien \surname{Guenneau}}
%\affiliation{Institut Fresnel-CNRS (UMR 6133), University of
%Aix-Marseille, 13397 Marseille cedex 20, France}
%\author{Stefan \surname{Enoch}}
%\affiliation{Institut Fresnel-CNRS (UMR 6133), University of
%Aix-Marseille, 13397 Marseille cedex 20, France}
%\date{\today}
%\date{July 29, 2009}

\author{Andr\'e Diatta,$^{1}$ Guillaume Dupont,$^{2}$\\ S\'ebastien Guenneau,$^{1,2}$
and Stefan Enoch$^{2}$}
%\address{$^{1}$Department of Mathematical Sciences, Peach Street, Liverpool L69 3BX, UK}
%\email{guenneau@liverpool.ac.uk}
%\address{$^{2}$Institut Fresnel, UMR CNRS 6133, Universit\'e Aix-Marseille III, 13397 Marseille, France}
\maketitle
\date{}
\noindent
{\footnotesize
\thanks{$^{1}$Department of Mathematical Sciences, Peach Street, Liverpool L69 3BX, UK\\ Email addresses: adiatta@liv.ac.uk; guenneau@liv.ac.uk
\\
$^{2}${\em{Institut Fresnel, UMR CNRS
6133, University of Aix-Marseille III,
\newline\noindent
case 162, F13397 Marseille Cedex 20, France. 
Email addresses:guillaume.dupont@fresnel.fr; sebastien.guenneau@fresnel.fr; stefan.enoch@fresnel.fr}};
\\

}
}

\begin{abstract}{\footnotesize
This paper extends the proposal of [Phys. Rev. Lett. 101, 203901-4
(2008)] to invisibility carpets for rigid planes and cylinders in
the context of pressure acoustic waves propagating in a compressible
fluid. Carpets under consideration here do not touch the ground:
they levitate in mid-air (or float in mid-water), which leads to
approximate cloaking for an object hidden underneath, or touching
either sides of a square cylinder on, or over, the ground. The
tentlike carpets attached to the sides of a square cylinder
illustrate how the notion of a carpet on a wall naturally
generalizes to sides of other small compact objects. We then extend
the concept of flying carpets to circular cylinders and show that
one can hide any type of defects under such circular carpets,  and
yet they still scatter waves just like a smaller cylinder on its
own. Interestingly, all these carpets are described by non-singular
acoustic parameters. To exemplify this important aspect, we propose
a multi-layered carpet consisting of isotropic homogeneous fluids
with constant bulk modulus and varying density which works over a
finite range of wavelengths.}
\end{abstract}

\noindent
{\footnotesize \bf \underline{Pacs:}} 00.00, 20.00, 42.10, 42.79.-e; 02.40.-k; 41.20.-q


\maketitle

%%%%%%%%%%%%%%%%%%%%%%%%%%%%%%%%%%%%%%%%%%%%%%%%%%%%%%%%%%%%%%%
%%%%%%%%%%%%%%%%%%% definition of symbols  %%%%%%%%%%%%%%%%%%%%
%%%%%%%%%%%%%%%%%%%%%%%%%%%%%%%%%%%%%%%%%%%%%%%%%%%%%%%%%%%%%%%
\def\index{\sigma}
\def\ep{\varepsilon}
\def\om{\omega}
\def\rot{\boldsymbol{\nabla} \times}
\def\div{\boldsymbol{\nabla} \cdot}
\def\p1{\partial_1}
\def\p2{\partial_2}
\def\p3{\partial_3}
\def\vE{\boldsymbol{E}}
\def\vH{\boldsymbol{H}}
\def\vF{\boldsymbol{F}}
\def\vG{\boldsymbol{G}}
\def\vJ{\boldsymbol{J}}
\def\vU{\boldsymbol{U}}
\def\vp{\boldsymbol{p}}
\def\ve{\boldsymbol{e}}
\def\vx{\boldsymbol{x}}
\def\vy{\boldsymbol{y}}
\def\vk{\boldsymbol{k}}
\def\F{\mathbb{F}}
\def\T{\mathbb{T}}
\def\N{\mathbb{N}}
\def\Z{\mathbb{Z}}
\def\R{\mathbb{R}}
\def\C{\mathbb{C}}

%%%%%%%%%%%%%%%%%%%%%%%%%%%%%%%%%%%%%%%%%%%%%%%%%%%%%%%%%%%%%%%
%%%%%%%%%%%%%%%%%%%      COMMENTAIRES      %%%%%%%%%%%%%%%%%%%%
%%%%%%%%%%%%%%%%%%%%%%%%%%%%%%%%%%%%%%%%%%%%%%%%%%%%%%%%%%%%%%%
%{\textbf{}
%%%%%%%%%%%%%%%%%%%%%%%%%%%%%%%%%%%%%%%%%%%%%%%%%%%%%%%%%%%%%%%
%%%%%%%%%%%%%%%%%%%        SECTION 1       %%%%%%%%%%%%%%%%%%%%
%%%%%%%%%%%%%%%%%%%%%%%%%%%%%%%%%%%%%%%%%%%%%%%%%%%%%%%%%%%%%%%

\section{Introduction}
There is currently a keen interest in electromagnetic metamaterials
within which very unusual phenomena such as negative refraction and
focussing effects involving the near field can occur
\cite{veselago,pendry_prl00,smith00,sar_rpp05}. A circular cylinder
coated with a negative refractive index displays anomalous
resonances \cite{Ross_cloaking} and can even cloak a set of dipoles
located in its close neighborhood \cite{graeme}. The dielectric
cylinder itself can be made transparent with a plasmonic coating
\cite{engheta}.

However, cloaking of arbitrarily sized objects requires anisotropic
heterogeneous media designed using transformation optics
\cite{pendrycloak,leonhardt}. The first experimental demonstration
of an invisibility cloak was obtained at $8.5$GHz and fueled the
interest in this new field of optics \cite{cloakex}. Electromagnetic
cloaks also allow for mirage effects \cite{zolla_oplett07} and their
efficiency very much depend upon the smoothness of their boundaries
\cite{diatta-guenneau}. Importantly, mathematicians have also
proposed some models of cloaks, but in the context of inverse
problems in tomography \cite{greenleaf,kohn} and also further
explained what are the appropriate boundary conditions on the inner
boundary of cloaks \cite{weder1,weder2}. These latter works open
new vistas in acoustic cloaks, and thus attract growing attention in
the physics and mathematics communities. Whereas cloaking of
pressure waves in two-dimensional \cite{cummernjp,sanchez} and
three-dimensional \cite{pendryprl,chen07} fluids, anti-plane shear
waves in cylindrical bodies \cite{njp2008}, and flexural waves in
thin-elastic plates \cite{prbihar,prlbihar} is well understood by
now, cloaking of in-plane coupled shear and pressure elastic waves
still remains elusive \cite{milton,brunapl} as the Navier equations
do not retain their form under geometric transforms. Such cloaks
might involve (complex) pentamode materials such as proposed in
\cite{norris}. However, it has been know for over ten years that
coated cylinders might become neutral in the elastostatic limit
\cite{bigoni98}, and this route might well be worthwhile pursuing
more actively with e.g. pre-stressed coated elastic cylinders, as
this would allow for a simple experimental setup.

During the same decade, some theoretical and experimental progress
has been made towards a better understanding of band spectra for
linear surface water waves propagating in arrays of rigid cylinders
\cite{feng06,prlchou,jfmchou,mciver,torres,hupre}, or over a bottom
with periodically drilled holes. Focussing effect of surface water
waves was also investigated by a handful of research groups in
arrays of circular and square cylindrical holes
\cite{hupre,farhat_pre}, as well as using metamaterials designed as
fluid networks \cite{fanglens}, \cite{page}. However, a further control of
surface water waves can be obtained via an alternative route, that
of transformation acoustics. It has been actually demonstrated that
broadband cloaking of surface water waves can be achieved with a
structured cloak, with an experimental confirmation at $10$
Hz\cite{farhat_prl}.

In this paper, we focus our analysis on cloaking of pressure waves
with carpets \cite{pendry-carpet} which have recently led to
experiments in the electromagnetic context
\cite{smithcarpet,gabrielli-carpet}. Here, we would like to render
e.g. pipelines lying at the bottom of the sea or floating in
mid-water undetectable for a boat sonar. These pipelines are
considered to be infinitely long straight cylinders with a
cross-section which is of circular or square shape. A pressure wave
incident from above (the surface of the sea) hits the pipeline, so
that the reflected wave reveals its presence to the sonar boat. We
would like to show that we can hide the pipeline under a cylindrical
carpet (a metafluid) so that the sonar only detects the wave
reflected by the bottom of the sea.

\section{Governing equations for pressure and transverse electric waves}
For an inviscid fluid with zero shear modulus, the linearized
equations of state for small amplitude perturbations from
conservation of momentum, conservation of mass, and linear
relationship between pressure and density are
\begin{eqnarray}
\begin{array}{l}
\displaystyle \rho_0 \frac{\partial {\bf v}}{\partial t} = -\nabla p
\cr \displaystyle \frac{\partial {p}}{\partial t} =
-\lambda\nabla\cdot{\bf v} \cr
\end{array}
\end{eqnarray}
where $p$ is the scalar pressure, ${\bf v}$ is the vector fluid
velocity, $\rho_0$ is the unperturbed fluid mass density (a mass in
kilograms per unit volume in meters cube), and $\lambda$ is the
fluid bulk modulus (i.e. it measures the substance's resistance to
uniform compression and is defined as the pressure increase needed
to cause a given relative decrease in volume, with physical unit in
Pascal). This set of equations admits the usual compressional wave
solutions in which fluid motion is parallel to the wavevector.

In cylindrical coordinates with $z$ invariance, and letting the mass
density be anisotropic but diagonal in these coordinates, the time
harmonic acoustic equations of state simplify to (the $\exp
(-j\omega t)$ convention is used throughout)
\begin{eqnarray}
\nabla\cdot \left(\rho_0^{-1} \nabla p\right) + \omega^2\lambda^{-1}
p = 0 \; , \label{govpressure}
\end{eqnarray}
where $\omega$ is the angular pressure wave frequency (measured in
radians per unit second). Importantly, this equation is supplied
with Neumann boundary conditions on the boundary of rigid defects
(no flow condition).

Cummer and Schurig have shown that this equation holds for
anisotropic heterogeneous fluids in cylindrical geometries
\cite{cummernjp}, and they have drawn some comparisons with
transverse electromagnetic waves. In the transverse electric
polarization (longitudinal magnetic field parallel to the cylinder's
axis):
\begin{eqnarray}
\nabla\cdot \left(\varepsilon_r^{-1} \nabla H_z\right) +
\omega^2\varepsilon_0\mu_0 H_z = 0 \; . \label{govmagz}
\end{eqnarray}
where $H_z$ is the longitudinal (only non-zero) component of the
magnetic field, $\varepsilon_r$ is the dielectric relative
permittivity, $\varepsilon_0\mu_0$ is the inverse of the square
velocity of light in vacuum, and $\omega$ is the angular transverse
electric wave frequency (measured in radians per unit second).
Importantly, this equation is supplied with Neumann boundary
conditions on the boundary of infinite conducting defects.

In this paper, we look at such `acoustic' models, for the case of
flying carpets which are associated with geometric transforms. While
we report computations for a pressure field, results apply mutatis
mutandis to transverse electric waves making the changes of
variables
\begin{eqnarray}
p\longmapsto H_z \; , \; \rho_0\longmapsto\varepsilon_r \; , \;
\lambda^{-1}\longmapsto\varepsilon_0\mu_0 \; ,
\label{correspondence}
\end{eqnarray}
in (\ref{govpressure}). However, after geometric transform,
(\ref{govpressure}) will involve an anisotropic (heterogeneous)
density $\underline{\underline{\rho}}$ and a varying (scalar) bulk
modulus $\lambda$, see (\ref{transfpressure}), whereas
(\ref{govmagz}) would involve an anisotropic (heterogeneous)
permittivity $\underline{\underline{\varepsilon}}$ and a varying
(scalar) permeability $\mu$, see (\ref{transfmaxwell}). It is
nevertheless possible to work with a reduced set of parameters, to
avoid a varying $\lambda$ in acoustics (resp. $\mu$ in optics), as
we shall see in the last section of the paper.

\section{From transformation optics to transformation acoustics}
Let us consider a map from a co-ordinate system $\{u,v,w\}$ to the
co-ordinate system $\{x,y,z\}$ given by the transformation
characterized by $x(u,v,w)$, $y(u,v,w)$ and $z(u,v,w)$.

This change of co-ordinates is characterized by the transformation
of the differentials through the Jacobian:

\begin{eqnarray}
\left(%
\begin{array}{c}
  dx \\
  dy \\
  dz \\
\end{array}%
\right) = \mathbf{J}_{xu}
\left(%
\begin{array}{c}
  du \\
  dv \\
  dw \\
\end{array}%
\right) \; , \hbox{with} \; \mathbf{J}_{xu}=
\frac{\partial(x,y,z)}{\partial(u,v,w)} \;.
%\end{array}
\end{eqnarray}

In electromagnetics, this change of coordinates amounts to replacing
the different materials (often homogeneous and isotropic, which
corresponds to the case of scalar piecewise constant permittivity
and permeability) by equivalent inhomogeneous anisotropic materials
described by a transformation matrix ${\bf T}$ (metric tensor). The
idea underpinning acoustic invisibility \cite{pendrycloak} is that
newly discovered metamaterials should enable control of the pressure
waves by mimicking the heterogeneous anisotropic nature of ${\bf T}$
with e.g. an anisotropic density, in a way similar to what was
recently achieved with the permeability and permeability tensors in
the microwave regime in the context of electromagnetism
\cite{cloakex}.

In the sequel, we propose an alternative derivation of the analogies
between transformation optics and acoustics first drawn by Cummer
and Schurig \cite{cummernjp}. Our proof involves a lemma first
established in \cite{fredseb} in the context of duality relations
for the Maxwell system in checkerboards. On a geometric point of
view, the matrix $\mathbf{T} \!= \! \mathbf{J}^T
\mathbf{J}/\det(\mathbf{J})$ is a representation of the metric
tensor. The only thing to do in the transformed coordinates is to
replace the materials (dielectric, homogeneous and isotropic) by
equivalent ones which now exhibit some magnetism, which are
heterogeneous (dependance upon $u,v,w$ co-ordinates) and
anisotropic. Their properties are given by \cite{zolla_oplett07}
\begin{eqnarray}
\underline{\underline{\epsilon'}} =\epsilon_r \mathbf{T}^{-1} \; ,
\quad
 \hbox{and} \quad
\underline{\underline{\mu'}}=\mathbf{T}^{-1} \; . \label{epsmuT}
\end{eqnarray}
In transverse electric polarisation, the Maxwell operator in the
transformed coordinates writes as
\begin{eqnarray}\label{eq:Hl}
\nabla\times\left(
\underline{\underline{\varepsilon'}}^{-1}\nabla\times {\bf H}_l
\right) - \mu_0\varepsilon_0\omega^2
\underline{\underline{\mu'}}{\bf H}_l= {\bf 0}
\end{eqnarray}
where ${\bf H}_l=H_z(x,y){\bf e}_z$, $
\underline{\underline{\varepsilon'}}$ and
$\underline{\underline{\mu'}}$ are defined by (\ref{epsmuT}).

We would like to deduce the expression of (\ref{govmagz}) in the
transformed coordinates from the vector equation (\ref{eq:Hl}). For
this, we need the following result:

%We note that there is no change in the wave-speed in the media which
%is actually defined by
%\begin{equation}
%{(\underline{\underline{\rho'}}\underline{\underline{\lambda'}}^{-1})}^{1/2}
%=\sqrt{\frac{\rho}{\lambda}} \mbox{Id}
%\end{equation} where $\mbox{Id}$ is the $3\times 3$ identity matrix.
%since the density $\rho$ and the bulk modulus $\lambda$ undergo the
%same transformation.

{\bf Property}: Let ${\bf M}$ be a real symmetric matrix defined as
follows
\begin{eqnarray}
{\bf M} = \left(
\begin{array}{ccc}
 m_{11} & m        & 0 \cr
 m      & m_{22}   & 0 \cr
 0      & 0        & m_{33}
\end{array}
\right) = \left(
\begin{array}{ccc}
 \tilde{{\bf M}} & 0 \cr
 0         & m_{33}
\end{array}
\right) \; .
\end{eqnarray}

Then we have
\begin{eqnarray}
\begin{array}{ll}
&\nabla\times \Biggl ( {\bf M} \nabla\times \Bigl ( u(x,y)
\mathbf{e}_z
\Bigr ) \Biggr ) \nonumber\\
&= - \nabla\cdot \Biggl ( \tilde{{\bf M}}^{-1}
\hbox{det}(\tilde{{\bf M}})\nabla u(x,y) \Biggr )\mathbf{e}_z
%\label{prop}
\end{array}
.
\end{eqnarray}

Indeed, we note that
\begin{eqnarray}
\begin{array}{ccc}
&\nabla\times \Biggl ( {\bf M} \nabla\times \Bigl ( u(x,y)
\mathbf{e}_z \Bigr ) \Biggr ) \nonumber\\
&= \displaystyle{- \Biggl ( \frac{\partial}{\partial x} \Bigl (
m_{22} \frac{\partial u}{
\partial x} - m \frac{\partial u}{\partial y} \Bigr )} \nonumber \\
& \displaystyle{+ \frac{\partial}{\partial y} \Bigl ( m_{11}
\frac{\partial u}{\partial y} - m \frac{\partial u}{\partial x} \Bigr ) \Biggr ) \mathbf{e}_z}
\end{array}.
\end{eqnarray}

Furthermore, let ${\bf M}'$ be defined as
\begin{eqnarray}
{\bf M}' = \left(
\begin{array}{ccc}
%\begin{pmatrix}
 m'_{11}      & m'_{12} \cr
 m'_{21}      & m'_{22}
%\end{pmatrix}
\end{array}
\right) \; .
\end{eqnarray}
We have $$\nabla\times \Bigl ( {\bf M} \nabla\times \Bigl ( u(x,y)
\mathbf{e}_z \Bigr ) \Biggr ) = - \nabla\cdot \Biggl ( {\bf M}'
\nabla u \Biggr )\mathbf{e}_z \; ,$$ if and only if
$$m'_{11} \frac{\partial u}{\partial x} + m'_{12} \frac{\partial u}{\partial y}
=m_{22} \frac{\partial u}{\partial x} - m \frac{\partial u}{\partial y}$$
$$m'_{21} \frac{\partial u}{\partial x} + m'_{22} \frac{\partial u}{\partial y}
=m_{11} \frac{\partial u}{\partial y} - m \frac{\partial u}{
\partial x} \; ,$$
which is true if ${\bf M}' = \tilde{{\bf
M}}^{-1}\hbox{det}(\tilde{{\bf M}})$.

\noindent Using the above property, from (\ref{eq:Hl}), we derive
that the transformed equation associated with (\ref{govmagz}) reads
\begin{eqnarray} \nabla\cdot
\underline{\underline{\epsilon'}}_T^{-1} \nabla H_z + \omega^2
\epsilon_0\mu_0 T_{zz}^{-1} H_z = 0 \; , \label{transfmaxwell}
\end{eqnarray}
with
\begin{eqnarray}\underline{\underline{\epsilon'}}_T^{-1}=\varepsilon_r^{-1}\tilde{{\bf
T}}/\hbox{det}({\tilde{\bf T}}). \end{eqnarray}
Here, $\tilde{{\bf T}}$ denotes
the upper diagonal part of the transformation matrix ${\bf T}$ and
$T_{zz}$ its third diagonal entry.

\noindent Invoking the one-to-one correspondence
(\ref{correspondence}), we infer that the transformed equation
associated with (\ref{govpressure}) reads
\begin{eqnarray} \nabla\cdot
\underline{\underline{\rho'}}_T^{-1} \nabla p + \omega^2
{\lambda}^{-1}T_{zz}^{-1} p = 0 \; , \label{transfpressure}
\end{eqnarray}
with \begin{eqnarray}\underline{\underline{\rho'}}_T^{-1}=\rho_0^{-1}\tilde{{\bf
T}}/\hbox{det}({\tilde{\bf T}}).\end{eqnarray}

In the sequel we will also consider a compound transformation. Let
us consider three coordinate systems $\{u,v,w\}$, $\{X,Y,Z\}$, and
$\{x,y,z\}$ (possibly on different regions of spaces).
The two successive changes of coordinates are given by the Jacobian
matrices $\mathbf{J}_{xX}$ and $\mathbf{J}_{Xu}$ so that
\begin{eqnarray}
\mathbf{J}_{xu}=\mathbf{J}_{xX}\mathbf{J}_{Xu} \; . \label{compjac}
\end{eqnarray}
This rule naturally applies for an arbitrary number of coordinate
systems.

\section{Flying carpets over a flat ground plane} 
\begin{figure}[ht]
{\includegraphics[width=8cm,angle=0] {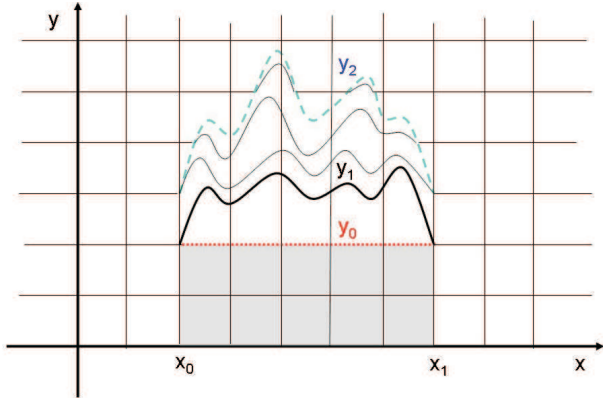}}
\mbox{}\vspace{-0.4cm}\caption{(Color online) Construction of a
carpet above the $x$-axis. The transformation
(\ref{eq:flyingcarpet}) shrinks the region between the two curves
$y={\bf y_0}$ (dotted red) and $y=y_2(x)$ (dashed blue) and two
vertical segments $x=x_0$ and $x=x_1$ into the region between the
curves $y=y_1(x)$ (solid black) and $y=y_2(x)$ (dashed blue) and the
vertical segments $x=x_0$ and $x=x_1$ (carpet). The curvilinear
metric inside the carpet is described by the transformation matrix
${\bf T}$, see (\ref{invtx}), corresponding to density and bulk
modulus of the metafluid given by (\ref{epsmuT}). The grey rectangle
can be either filled with air/ambient fluid (flying/floating carpet)
or be replaced by a rigid cylinder.} \label{fig:flyingcarpetdraw}
\end{figure}

This
section is dedicated to the study of carpets levitating  above a
ground plane, that conceal to certain extent any object placed
anywhere underneath them from plane waves incident from above. The
construction of the carpet is a generalization of those considered
in \cite{pendry-carpet} to carpets flying over ground planes or
located on either sides of rectangular objects, as shown in Fig.
\ref{fig:flyingcarpetdraw} with certain altitude $y=y_0$. In the
context of pressure waves, this corresponds for instance to the
physical situation of a carpet which flies in mid-air if $y=0$ is
the altitude of the ground, or a carpet which floats in mid-waters
if $y=0$ stands for the bottom of the sea. This formalism allows us
to study carpets which are either flying/floating on their own, or
which are touching a cylindrical object on the ground or in
mid-air/water.

\subsection{The construction of carpets over horizontal planes}
When the carpet is made on the y-axis (we call it a carpet flying above the
x-axis), we consider a transformation mapping the region enclosed
between two curves $(x,y_0)$  and  $(x,y_2(x))$ to the one comprised
between $(x,y_1(x))$  and  $(x,y_2(x))$ as in Fig.
\ref{fig:flyingcarpetdraw}, where $(x,y_0)$  is mapped on
$(x,y_1(x))$ and  $(x,y_2(x))$ is fixed point-wise, of the form
\begin{eqnarray}\label{eq:flyingcarpet}
\left\{
\begin{array}{lr}
x'=x\\
y'=\alpha (x)y+ \beta(x) \text{~with~} \alpha= \frac{y_2-y_1}{y_2-y_0} ~\text{and} ~
\beta= \frac{y_1-y_0}{y_2-y_0}y_2\\
z'=z
\end{array}
\right.
\end{eqnarray}
So the inverse of this transformation is given by
\ \begin{eqnarray}\label{eq:inversetransfo}
\left\{
\begin{array}{lr}
x=x'\\
y=\frac{y'-\beta(x')}{\alpha (x')}\\
z=z'
\end{array}
\right.
 \end{eqnarray}
 The above transformation (\ref{eq:inversetransfo}) has the following Jacobian matrix
   \begin{eqnarray}
\mathbf{J}_{xx'}=\frac{\partial (x,y,z)}{\partial (x',y',z')}=\begin{pmatrix}1 &0&0\\
g & \frac{1}{\alpha} & 0\\
0& 0 & 1 \end{pmatrix}.
 \end{eqnarray}
in which we have set
 \begin{eqnarray}
g&:=&\frac{\partial y }{\partial x'} =\frac{1}{\alpha^2}
\left(-\alpha\frac{d\beta}{dx'}-(y'-\beta)\frac{d\alpha}{dx'}\right)
\nonumber\\
 &=&-\frac{(y_2-y')(y_2-y_0)}{(y_2-y_1)^2}\frac{dy_1}{dx}\nonumber\\
&+&\frac{(y_1-y')(y_1-y_0)
}{(y_2-y_1)^2}\frac{dy_2}{dx}.
 \end{eqnarray}
 Hence we get
  \begin{eqnarray}
\mathbf{T}^{-1}=\mathbf{J}_{xx'}^{-1}\mathbf{J}_{xx'}^{-T}\det(\mathbf{J}_{xx'})=
\begin{pmatrix}\frac{1}{\alpha} & - g &0\\
- g & (1+g^2) \alpha & 0\\
0& 0 & \frac{1}{\alpha} \end{pmatrix}. \label{invtx}
 \end{eqnarray}

\subsection{Analysis of the metamaterial properties}\label{chap:IVB}
Let us now look at an interesting feature of the invisibility
carpet. The eigenvalues of ${\bf T}^{-1}$ are given by:

\begin{eqnarray}
i=1,2 \; : \;
\lambda_{i}&=&
\frac{1}{2\alpha}\left( 1+\alpha^2+g^2
\alpha^2 
+ (-1)^{i}\sqrt{-4\alpha^2+ (1+\alpha^2+g^2 \alpha^2)^2}\right), 
\nonumber
\\ \lambda_3&=&\displaystyle{\frac{1}{\alpha}}.
\end{eqnarray}

We note that $\lambda_i$, $i=1,2$, and $\lambda_3$ are strictly
positive functions as obviously $1+\alpha^2+g^2
\alpha^2>\sqrt{-4\alpha^2+ \left(1+\alpha^2+g^2 \alpha^2 \right)^2}$
and also $\alpha
>0$. This establishes that ${\bf T}^{-1}$ is not a singular matrix for a
two-dimensional carpet even in the case of curved ground planes,
which is one of the main advantages of carpets over cloaks
\cite{pendry-carpet}. The broadband nature of such carpets remains
to be investigated when one tries to mimic their ideal material
parameters with structured media.

 The carpets in Fig. \ref{fig2} and \ref{fig3} are made of a semi-circle
  $y=y_2(x)$ as an outer curve and a semi-ellipse $y=y_1(x)$ as its inner curve,
   both centered at $(a_0,b_1)$ with
\begin{eqnarray}y_0&=&b_0, ~~ y_1(x)=b_1+(1-\frac{k_0}{r_0})
\sqrt{r_0^2-(x-a_0)^2},\nonumber\\ y_2(x)&=&b_1+\sqrt{r_0^2-(x-a_0)^2}.
\end{eqnarray}
Plugging the numerical values $r_0:=0.2, k_0=0.1, a_0=0$ in, one
gets
%$
\begin{eqnarray}g=\frac{(5x(50b_1(b_0-b_1)+1-25x^2+50y(b_1-b_0))}{(1-25x^2)^{\frac{3}{2}}},\nonumber\end{eqnarray}%$,
where $b_0$ is the $y$-coordinate of the ground (typically, the
ground can be taken as $y=0$, so that $b_0=0$) and $b_1$ is the
hight, measured on the $y$-axis from the origin, at which the carpet
is flying.

\subsection{The construction of carpets over vertical planes}
We note that for a carpet flying above the $y$-axis, if we consider
the transformation mapping the region enclosed between two curves
$(x_0(y),y)$  and  $(x_2(y),y)$ to the one comprised between
$(x_1(y),y)$  and  $(x_2(y),y)$ as in Fig.
\ref{fig:flyingcarpetdraw}, where again $(x_0(y),y)$  is mapped on
$(x_1(y),y)$ and  $(x_2(y),y)$ is fixed point-wise, that is, if
$$
\left\{
\begin{array}{lr}
x'=\alpha (y)x+ \beta(y) \text{~with~} \alpha = \frac{x_2-x_1}{x_2-x_0} ~\text{and}~
\beta= \frac{x_1-x_0}{x_2-x_0}x_2\\
y'=y\\
z'=z
\end{array}
\right.
$$
then the Jacobian of the inverse transformation now reads
 \begin{eqnarray}
\mathbf{J}_{xx'}=\frac{\partial(x,y,z)}{\partial (x',y',z')}=\begin{pmatrix} \frac{1 }{\alpha} &h&0\\
0 & 1 & 0\\
0& 0 & 1 \end{pmatrix}
\end{eqnarray} where $h$ is defined as \begin{eqnarray}
h&:=&\frac{\partial x }{\partial y'} =\frac{1}{\alpha^2}\left(-\alpha
\frac{d\beta}{dy'}-(x'-\beta)\frac{d\alpha}{dy'}\right)
\nonumber\\
 &=&-\frac{(x_2-x')(x_2-x_0)}{(x_2-x_1)^2}\frac{dx_1}{dy} \nonumber\\
&+& \frac{(x_1-x')(x_1-x_0)
}{(x_2-x_1)^2}\frac{dx_2}{dy}.
 \end{eqnarray}
 and hence we get
  \begin{eqnarray}
\mathbf{T}^{-1}
=\begin{pmatrix}(1+h^2) \alpha & - h &0\\
- h & \frac{1}{\alpha} & 0\\
0& 0 & \frac{1}{\alpha} \end{pmatrix}. \label{invty}
 \end{eqnarray}

The same analysis as in Section \ref{chap:IVB} literally applies here, as well.

\subsection{Numerical results for a carpet over a plane}
\begin{figure}[ht]
\scalebox{0.4}
{\includegraphics{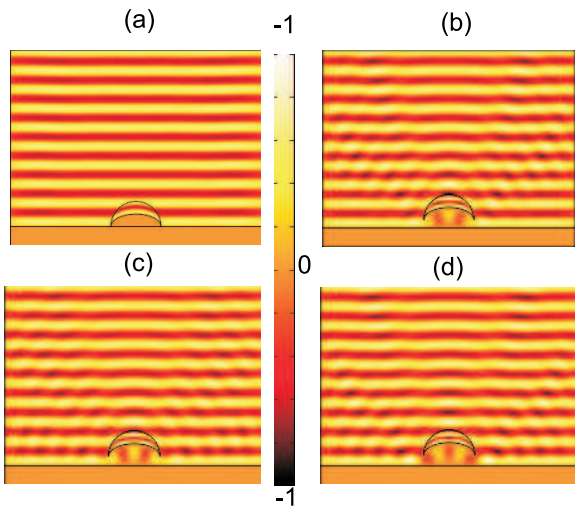}}
\mbox{}\vspace{-0.0cm}\caption{(Color online) 2D plot of the real
part of the total pressure field $\Re{e}(p)$: Scattering by a
pressure plane wave of wavelength $0.15$ incident from the top on a
flat ground plane with a carpet above it, here $y_1(x)=
b_1+1/2\sqrt{0.04-x^2}$ and $y_2(x)= b_1+\sqrt{0.04-x^2}$. (a)
carpet touching the ground, (b) carpet flying at altitude $b_1=0.7$,
(c) at $b_1=0.9$ and (d) at $b_1=1$. The optimal altitude (for a
flying carpet) $b_1=0.9$ is noted.} \label{fig2}
\end{figure}

We first look at the case of a pressure plane wave incident upon a
flat rigid ground plane (Neumann boundary conditions) and a carpet
above it. The inner boundary of the carpet is rigid (Neumann
boundary conditions). We report these results in Figure \ref{fig2}
where we can see that the altitude $y'=0.9$ leads to less scattering
than the other two flying carpets. Of course, the carpet attached to
the ground plane leads to perfect invisibility.

\begin{figure}[ht]
\scalebox{0.4}
{\includegraphics{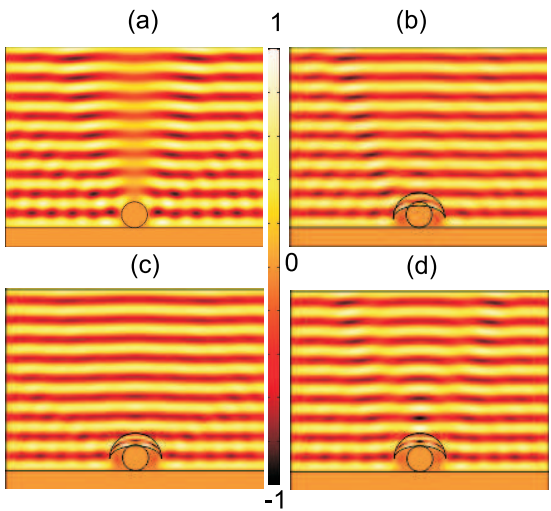}}
\mbox{}\vspace{-0.0cm}\caption{(Color online) 2D plot of the real
part of the total pressure field $\Re{e}(p)$: Scattering by a
pressure plane wave of wavelength $0.15$ incident from above on a
circular object of radius $r=0.6$ lying on a flat ground plane; (a)
obstacle without carpet, (b) with a carpet flying at altitude
$b_1=0.7$, (c) at $b_1=0.9$ and (d) at $b_1=1$. The optimal altitude
$b_1=0.9$ is noted.} \label{fig3}
\end{figure}

We then look at the case of the rigid ground plane with a rigid
circular obstacle on top of it. Some Neumann boundary conditions are
set on the ground plane, the inner boundary of the carpet and the
rigid obstacle. Once again, we can see in Figure \ref{fig3} that the
altitude $y'=0.9$ for the flying carpet is the optimal one.

\subsection{Numerical results for a carpet over a square cylinder
over a plane}

In this paragraph, we would like to render a pipeline lying on the
bottom of the sea or floating in mid-water undetectable for a boat
with a sonar at rest just above it on the surface of the sea.
However, instead of reducing its scattering cross-section like in
acoustic cloaks, we rather mimic that of another obstacle, say a
square rigid cylinder.

\begin{figure}[ht]
\scalebox{0.4}{\includegraphics{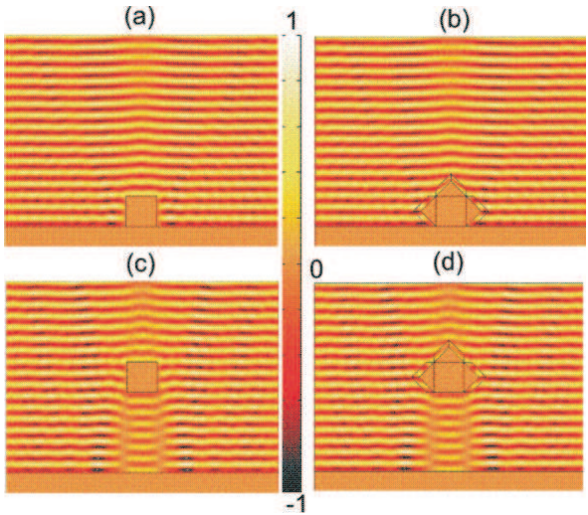}}
\mbox{}\vspace{-0.2cm}\caption{(Color online) 2D plot of the real
part of the total pressure field $\Re{e}(p)$: Scattering by a
pressure plane wave of wavelength $0.15$ incident from above on a
rigid square cylinder of sidelength $d=0.4$ lying on a flat ground
plane on its own (a); with three (tentlike) carpets on it sides (b);
flying over the ground plane on its own (c); with three (tentlike)
carpets on its sides (d).} \label{fig4} \end{figure}

In Fig. \ref{fig4}, we plot the total field when the pressure
wave of wavelength $0.15$ is incident from above on a rigid square
cylinder of sidelength $0.4$. In panel (a) the obstacle touches the
ground (e.g. bottom of the sea) at $y=-1.26$ and is centered
about $x=0$. In panel (b), it has three carpets shaped
as tents attached to its sides. The construction of the carpets is as follows: The
right-most tent is defined by $x_0=0.2$, $x_1(y)=-y-0.66$,
$x_2(y)=-y-0.56$ (upper part) and $x_1(y)=y+1.46$, $x_2(y)=y+1.56$
(lower part). The left-most tent is defined by $x_0= - 0.2$,
$x_1(y)=-y-1.46$, $x_2(y)=-y-1.56$ (lower part), and $x_1(y)=y+0.66$,
$x_2(y)=y+0.56$ (upper part). The pressure and bulk modulus of these
left-most and right-most tents is deduced from the expression of the
transformation matrix for vertical walls, see (\ref{invty}).
Finally, the uppermost tent is defined by $y_0= - 0.86$,
$y_1(x)=-x-0.66$, $y_2(x)=-x-0.56$ (right part) and $y_1(x)=x-0.66$, $y_2(x)=x-0.56$ (to the left).
Here, we use the expression of the transformation matrix for
horizontal walls, see (\ref{invtx}).

We now look at the case of a floating rigid square cylinder of sidelength $0.4$ in
mid-water (e.g. a pipeline). In panel (c), the obstacle is flying on its own. In panel (d), it has three tentlike carpets on its sides and yet still scatters the incoming wave as in (c). The construction of the carpets in panel (d) is as
follows: The right-most tent is defined by $x_0=0.2$,
$x_1(y)=-y+0.4$, $x_2(y)=-y+0.5$ (upper part) and $x_1(y)=y+0.4$,
$x_2(y)=y+0.5$ (lower part). Similarly, the left-most tent is defined by $x_0= -
0.2$, $x_1(y)=-y-0.4$, $x_2(y)=-y-0.5$ (lower part) and $x_1(y)=y-0.4$,
$x_2(y)=y-0.5$ (upper part). Finally, the uppermost tent is defined by $y_0=0.2$,
$y_1(x)=-x+0.4$, $y_2(x)=-x+0.5$ (right) and $y_1(x)=x+0.4$, $y_2(x)=x+0.5$(left).

We note that while forward and backward scattering in panels (b) and
(d) are not negligible (as would be the case for acoustic cloaks),
these are instantly recognizable as that of a square obstacle on the
ground (panel a) or floating in mid-water (panel c). The next
question to address is whether such a generalized cloaking works at
other incidences.

\subsection{Numerical results for a carpet over a square cylinder over a plane in grazing incidence}

We now look at the case of grazing incidence, keeping otherwise the
same configuration as in Fig. \ref{fig4}.  We note that the total
field for the square obstacle is exactly the same with and without
the three carpets, which is a further evidence that carpets allow
for multi-incidence cloaking. Of course, one can hide any object
(e.g. a semi-disc) under the carpets in panel (b) and this will
scatter as a square. However, in the case of a flying object, it is
required that the lower boundary of the hidden object be flat (e.g.
a semi-disc) in order to mimic the diffraction pattern associated
with a square obstacle.

Taking into account that it should be possible to design broadband
carpets as their material parameters are non-singular, our numerics
suggest that carpets are thus a very interesting alternative to
invisibility cloaks. They can either reduce the scattering cross
section of a rigid object or mimic that of another rigid object.

\begin{figure}[ht]
\scalebox{0.4}{\includegraphics{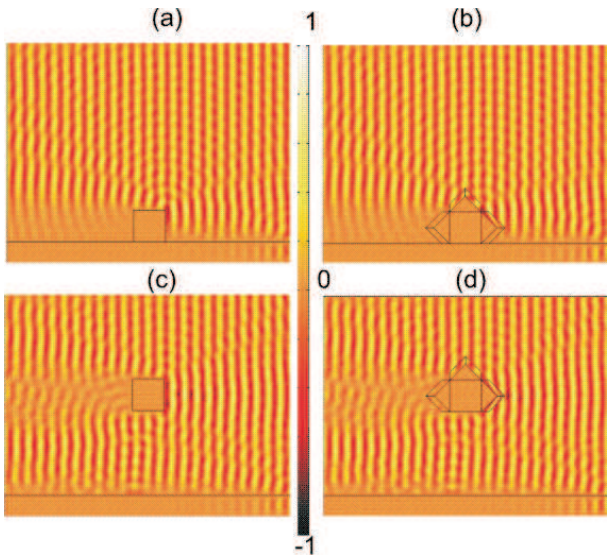}}
\mbox{}\vspace{-0.2cm}\caption{(Color online) 2D plot of the real
part of the total pressure field $\Re{e}(p)$: Scattering by a
pressure plane wave of wavelength $0.15$ incident from the right
(grazing incidence) on a rigid square cylinder of sidelength $d=0.4$
lying on a flat ground plane on its own (a); with three tentlike
carpets on it sides (b); flying over the ground plane on its own
(c); with three tentlike carpets on its sides (d).} \label{fig7}
\end{figure}

Finally, we should emphasize that we numerically checked that our
proposal for generalized carpets works fine for a pressure plane
wave incident on the square at any angles ranging from normal to
grazing incidence. This should not come as a surprise as the same
holds true in the original proposal by Li and Pendry
\cite{pendry-carpet}.

\section{Flying carpets surrounding a circular cylinder over a plane}
We propose in this section to design a carpet
which geometrically looks like Pendry's cylindrical cloak
\cite{pendrycloak}, but in fact, it scatters any incoming plane wave
just like a rigid cylinder of smaller radius $R_0$. The radius of
such an effective rigid cylinder can be of any size smaller than
that of the coated region. Such a carpet acts in a way similar to
the invisibility cloaks proposed by Greenleaf et al. and Kohn et al.
which are based upon the blow up of a small ball (of radius $\eta
\ll 1$) rather than a point, thereby leading to approximate
invisibility. However, in our case, the radius of this ball (a disc
in 2D) is finite, and our claim is that the carpet mimics the
electromagnetic response of the rigid circular cylinder, just like
previous carpets did for planes. The circular carpet actually plays
the opposite role to the super-scatterer of Nicorovici, McPhedran
and Milton whereby a cylinder surrounded by a coating of negative
refractive index material scatters as a cylinder of diameter larger
than the coating itself \cite{Ross_cloaking}.

However, whether this coated region is completely empty (e.g. Fig
\ref{fig:cylindriccarpet}(b) ) or many objects are hidden inside it
(see e.g. Fig. \ref{fig:cylindriccarpet}(d) where the small cylinder
of radius $R_0$ has actually been put back inside the carpet, along
with other objects), such a carpet produces a mirage effect that
tricks an external observer into believing that this whole region is
just the small rigid cylinder.

\subsection{The construction of circular carpets}
The carpet consists of a cylindrical region $C(\underline 0, R_1)$ of radius $R_1$ to be coated
 and the coating itself which is the space between an inner cylinder of radius $R_1$ and an
outer one of radius $R_2>R_1$. As above, the material properties of
this coating will be deduced by pullback, via a transformation that
fixes angles just like in \cite{pendrycloak}. The transformation in
\cite{pendrycloak} is indeed the particular case, of the one under
consideration here, associated with the value $R_0=0$ of the radius
of an imaginary small cylinder as explained below.

\begin{figure}[ht]
\scalebox{0.5}{\includegraphics{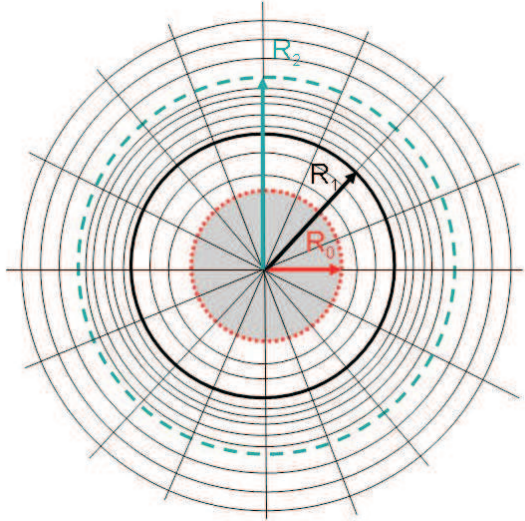}}
\mbox{}\vspace{-0.4cm}\caption{(Color online) Construction of a
circular carpet of inner radius $R_1$ (solid dark) and outer radius
$R_2$ (dashed blue) from cylinder $\mathcal S(\underline 0, R_0)$ of
smaller radius $R_0$ (dotted red). The transformation
(\ref{eq:circular}) shrinks the whole hollow cylindrical  region
$\mathcal A(R_0,R_2)$ of inner radius $R_0$ and outer radius $R_2$
into  its subset $\mathcal A(R_1,R_2)$, the cylinder $\mathcal
S(\underline 0, R_0)$ being stretched to $\mathcal S(\underline 0,
R_1)$ whereas $\mathcal S(\underline 0, R_2)$ is fixed point-wise.
Such a carpet, a metafluid with bulk modulus and density given by
(\ref{invtcir}) and (\ref{epsmuT}), scatters pressure waves as the
rigid cylinder $\mathcal S(\underline 0, R_0)$.}
\label{fig:carpetdraw-circ}
\end{figure}

\begin{figure}[ht]
\scalebox{0.4}{\includegraphics{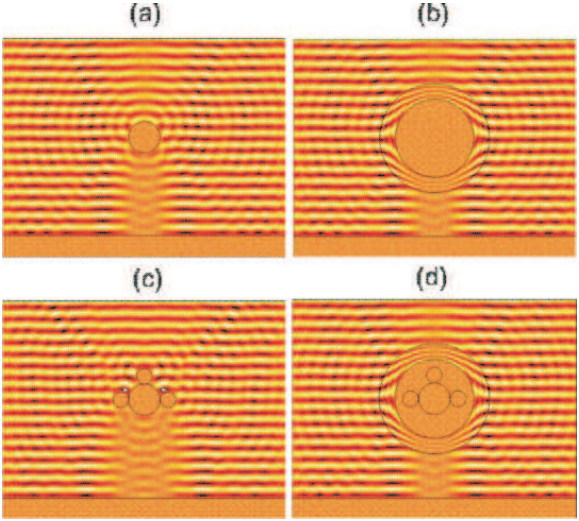}} 
\mbox{}\vspace{-0.2cm}\caption{(Color online) 2D plot of the real
part of the total pressure field $\Re{e}(p)$: Scattering by a plane
wave of wavelength $0.15$ incident from above on a flat ground plane
and (a) a circular object of radius $R_0=0.2$ flying at altitude on
its own; (b) an empty cylindrical region of radius $R_1=0.5$, the
"coated" region, surrounded by a carpet of inner radius $R_1$ and
outer radius $R_2=0.7$. This hollow cylindrical carpet is designed
to behave exactly like the object in (a) alone,  irrespective of the
form of any other additional object that may be enclosed inside; (c)
same object as in (a) with now three additional small rigid
cylinders touching it. The scattering is clearly different from that
in (a); (d) now the three objects in as (c) have been hidden inside
the carpet (b)  and yet an outer observer will not be able to tell
the scattering in (a) and in (d) apart.} \label{fig:cylindriccarpet}
\end{figure}

The carpet constructed below is a generalization of the flying
carpet studied in the previous section. The geometric transformation
we consider here, is a smooth diffeomorphism of the closure of the
outside $\mathbb R^3\backslash C(\underline 0, R_0)$  of a solid
cylinder $C(\underline 0, R_0)$ of radius $R_0,$ where $R_0<R_1.$ It
coincides with the identity map outside the solid cylinder
$C(\underline 0, R_2),$ fixing its boundary point-wise, but now maps
the region $\mathcal A(R_0,R_2)$  between the two coaxial cylinders
of respective radii  $R_0$ and $R_2$ into the space $\mathcal
A(R_1,R_2)$ between the cylinders of radii $R_1$ and $R_2,$ as in
Fig. \ref{fig:carpetdraw-circ}.
 More precisely, in $\mathcal A(R_0,R_2)$ this geometric transformation can be expressed as
\begin{eqnarray}\label{eq:circular}
\left\{
\begin{array}{lr}
r'=R_1+\alpha (r-R_0) \text{ ~with ~} \alpha= \frac{R_2-R_1}{R_2-R_0}\\
\theta'=\theta\\
z'=z
\end{array}
\right.
\end{eqnarray}
with inverse
$$
\left\{
\begin{array}{lr}
r=R_0+\frac{1}{\alpha} (r'-R_1)\\
\theta=\theta'\\
z=z'
\end{array}
\right.
$$

The Jacobian $\mathbf{J}_{rr'}$ of the latter is

\begin{eqnarray}
\mathbf{J}_{rr'}=\frac{\partial(r,\theta,z)}{\partial(r',\theta',z')}=\mbox{diag}(
\frac{1 }{\alpha},1,1).
\end{eqnarray}

Let us denote by $\mathbf{J}_{xr}$ the Jacobian of the change
$(r,\theta,z)\mapsto (x,y,z)$ from Cartesian to polar coordinates
and $\mathbf{J}_{rx}:=\mathbf{J}_{xr}^{-1}.$  The Jacobian
$\mathbf{J}_{xx'}$ of the above transformation in Cartesian
coordinates $(x',y',z') \mapsto (x,y,z)$ is obtained by applying the
chain rule (\ref{compjac}) to get
$\mathbf{J}_{xx'}=\mathbf{J}_{xr}\mathbf{J}_{rr'}\mathbf{J}_{r'x'},$
so that the tensor $\mathbf{T}^{-1}=\mathbf{J}_{xx'}^{-1}\mathbf{J}_{xx'}^{-T}\det(\mathbf{J}_{xx'})$
reads
\begin{eqnarray}
\mathbf{T}^{-1} &=&\begin{pmatrix}\frac{1+(m^2-1)\cos^2(\theta)}{m}&\frac{(m^2-1)\sin(\theta)
\cos(\theta)}{m}
&0\\
\frac{(m^2-1)\sin(\theta)\cos(\theta)}{m} &\frac{m^2+(1-m^2)\cos^2(\theta)}{m}&0\\0 &0
&\frac{m}{\alpha^2}
\end{pmatrix}\nonumber\\
&=&R(\theta)~\mbox{diag}(m,
\frac{1}{m},\frac{m}{\alpha^2})~R(-\theta), \label{invtcir}
\end{eqnarray}
where $
\mathbf{R}(\theta)
=\begin{pmatrix}\cos(\theta) & -\sin(\theta)  &0\\
\sin(\theta) & \cos(\theta) & 0\\
0& 0 & 1 \end{pmatrix}
$ is the matrix of the rotation with angle $\theta$ in the $xy$-plane and
  $m=\frac{\alpha r}{r'}=1-\frac{R_1-R_0}{(R_2-R_0)}\frac{R_2}{r'}.$

\subsection{Analysis of the metamaterial properties}
The tensor $\mathbf{T}^{-1}$ diagonalizes as $\mbox{diag}(m,
\frac{1}{m}, \frac{m}{\alpha^2})$ where
$m=1-\frac{R_1-R_0}{(R_2-R_0)}\frac{R_2}{r'}$ is bounded from below
and above as it satisfies $0<\alpha\frac{R_0}{R_1}\leq m \leq \alpha
.$ This means that the material parameters are not singular, unlike
the cloaking case as in \cite{pendrycloak}.

Obviously, one realizes that, when $R_0$ tends to zero, one recovers
the case in \cite{pendrycloak} where the material properties are no
longer bounded, but one of them tends to zero whilst another one
recedes to infinity, as we approach the inner boundary of the coated
region, see also \cite{kohn}.

\subsection{Mirage effect for a cylinder surrounded by a carpet}
We report the results of our simulations in Fig.
\ref{fig:cylindriccarpet} for a circular pipeline which is floating
in mid-water, see panel (a). We then replace this pipeline by a
circular carpet, see panel (b), which reflects a pressure plane wave
from above at wavelength $0.5$ in exactly the same way. When we add
three small pipelines to the original one, see panel (c), the
reflected field is obviously much different. However, when we
surround the four pipelines by the circular carpet, see panel (d),
the reflected wave is that of the original pipeline. Such a mirage
effect, whereby a rigid obstacle hides other ones in its
neighborhood, can thus be used for 'sonar illusions'. For
instance, an oil pipeline might reflect pressure waves like a coral
barrier so that a sonar boat won't catch its presence. Unlike for
earlier proposals of approximate cloaks \cite{greenleaf,kohn}
scattering waves like a small highly conducting object, we emphasize
here that we start the construction of the circular carpet by a
finite size disc.

\section{Multilayered circular carpet for broadband mirage effect}

\subsection{Reduced material parameters for circular carpets}
We now want to simplify the expression of the inverse of the
transformation matrix ${\bf T}$ in order to avoid a varying (scalar)
density (resp. permeability in optics). For this, we introduce the
reduced matrix $\mathbf{T}^{-1}_{red}=\mbox{diag}(\alpha,
\frac{\alpha}{m^2}, \frac{1}{\alpha})$ which amounts to multiplying
$\mathbf{T}^{-1}$ in (\ref{invtcir}) by $\alpha/m$. We deduce from
(\ref{govpressure}) the transformed governing equation for pressure
waves in reduced coordinates:
\begin{eqnarray} \nabla_{r,\theta}\cdot
\mbox{diag}(\frac{1}{\alpha}, \frac{m^2}{\alpha})\nabla_{r,\theta} p
+ \omega^2 {\lambda}^{-1}\alpha p = 0 \; , \label{transfpressurered}
\end{eqnarray}
and similarly for transverse electric waves using (\ref{govmagz}):
\begin{eqnarray} \nabla_{r,\theta}\cdot
\mbox{diag}(\frac{1}{\alpha}, \frac{m^2}{\alpha}) \nabla_{r,\theta}
H_z + \omega^2 \epsilon_0\mu_0 \alpha H_z = 0 \; ,
\label{transfmaxwellred}
\end{eqnarray}
where $m=1-\frac{R_1-R_0}{(R_2-R_0)}\frac{R_2}{r'}$ and
$\alpha=\frac{R_2-R_1}{R_2-R_0}$.

\subsection{Homogenized governing equations for optics and acoustics}
To illustrate our paper with a practical example, we finally choose
to design a circular carpet using 40 layers of isotropic homogeneous
fluids. These fluids have constant bulk modulus and varying density.
We report these computations in figure \ref{fig:multilayer}.

It is indeed well known that the homogenized acoustic equation for
such a configuration takes the following form:
\begin{eqnarray}
\nabla_{r,\theta}.
(\rho_0^{-1}\underline{\underline{\rho}}^{-1}\,\nabla_{r,\theta}
H_z) +\omega^2<\lambda^{-1}>\,p=0 \; ,\label{bihahom}
\end{eqnarray}
where $<\lambda^{-1}>=\int_0^1\lambda^{-1}(r)\, dr$ and with
$\underline{\underline{\rho}}$ a homogenized rank-2 diagonal tensor
(an anisotropic density) $\underline{\underline{\rho}}={\rm
Diag}(\rho_r,\rho_\theta)$ given by
\begin{eqnarray}
\underline{\underline{\rho}}={\rm Diag}({<\rho^{-1}>}^{-1},<\rho>)
\; .\label{parameter}
\end{eqnarray}

We note that if the cloak consists of an alternation of two
homogeneous isotropic layers of fluids of thicknesses $d_A$ and
$d_B$, with bulk moduli $\lambda_A$ and $\lambda_B$ and densities
$\rho_A$ and $\rho_B$, we have
\begin{eqnarray}
\begin{array}{lll}
&\displaystyle{\frac{1}{\rho_r}}=\displaystyle{\frac{1}{1+\eta}
\left(\frac{1}{\rho_A}+\frac{\eta}{\rho_B}\right)}
\; , \nonumber \\
& \nonumber \\
&\rho_\theta=\displaystyle{\frac{\rho_A+\eta \rho_B}{1+\eta}} \; ,
\; \displaystyle{<\lambda^{-1}}>=\displaystyle{\frac{1}{1+\eta}
\left(\frac{1}{\lambda_A}+\frac{\eta}{\lambda_B}\right)} \; ,
\end{array}
\end{eqnarray}
where $\eta=d_B/d_A$ is the ratio of thicknesses for layers $A$ and
$B$ and $d_A+d_B=1$.

Using the change of variables
\begin{eqnarray}
\nabla_{r,\theta}.
(\varepsilon_r^{-1}\underline{\underline{\varepsilon}}^{-1}\,\nabla_{r,\theta}
p)+\varepsilon_0\mu_0\omega^2\,H_z=0 \; ,\label{bihahom}
\end{eqnarray}
where $\underline{\underline{\varepsilon}}$ is a homogenized rank-2
diagonal tensor (an anisotropic permittivity)
$\underline{\underline{\varepsilon}}={\rm
Diag}(\varepsilon_r,\varepsilon_\theta)$ given by
\begin{eqnarray}
\underline{\underline{\varepsilon}}={\rm
Diag}({<\varepsilon^{-1}>}^{-1},<\varepsilon>) \; .\label{parameter}
\end{eqnarray}

\subsection{Acoustic paradigm: Reduced scattering with larger scatterer}
\begin{figure}[ht]
\scalebox{0.43}{\includegraphics{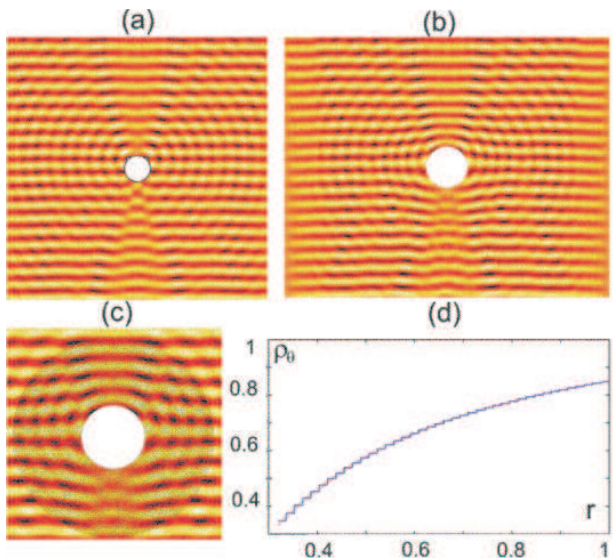}} %\includegraphics{cape2d1bis.eps}
\mbox{}\vspace{-0.3cm}\caption{(Color online) 2D plot of the real part
of the total pressure field $\Re{e}(p)$: Scattering by a plane wave
of wavelength $0.2$ incident from above on a flat ground plane and
(a) a circular object of radius $R_0=0.2$ flying at altitude on its
own; (b) an empty cylindrical region of radius $R_1=0.32$, the
"coated" region consisting of 40 layers of isotropic homogeneous
fluid, see closer view in (c), of constant bulk modulus and density
$\rho$, given in (d), surrounded by a carpet of inner radius $R_1$
and outer radius $R_2=1$. The red curves represents the variation of
$\rho_\theta=m^2/\alpha$ with respect to $r\in[0.32;1]$. The
piecewise constant blue curve is a staircase approximation of the
red curve, considering an alternation of 40 layers of density
$\rho_A\in [0.1890 ; 0.5493]$ and $\rho_B\in [1.7987 ; 2.1472]$
using the homogenized formula (\ref{parameter}).}
\label{fig:multilayer}
\end{figure}

\begin{figure}[ht]
\scalebox{0.47}{\includegraphics{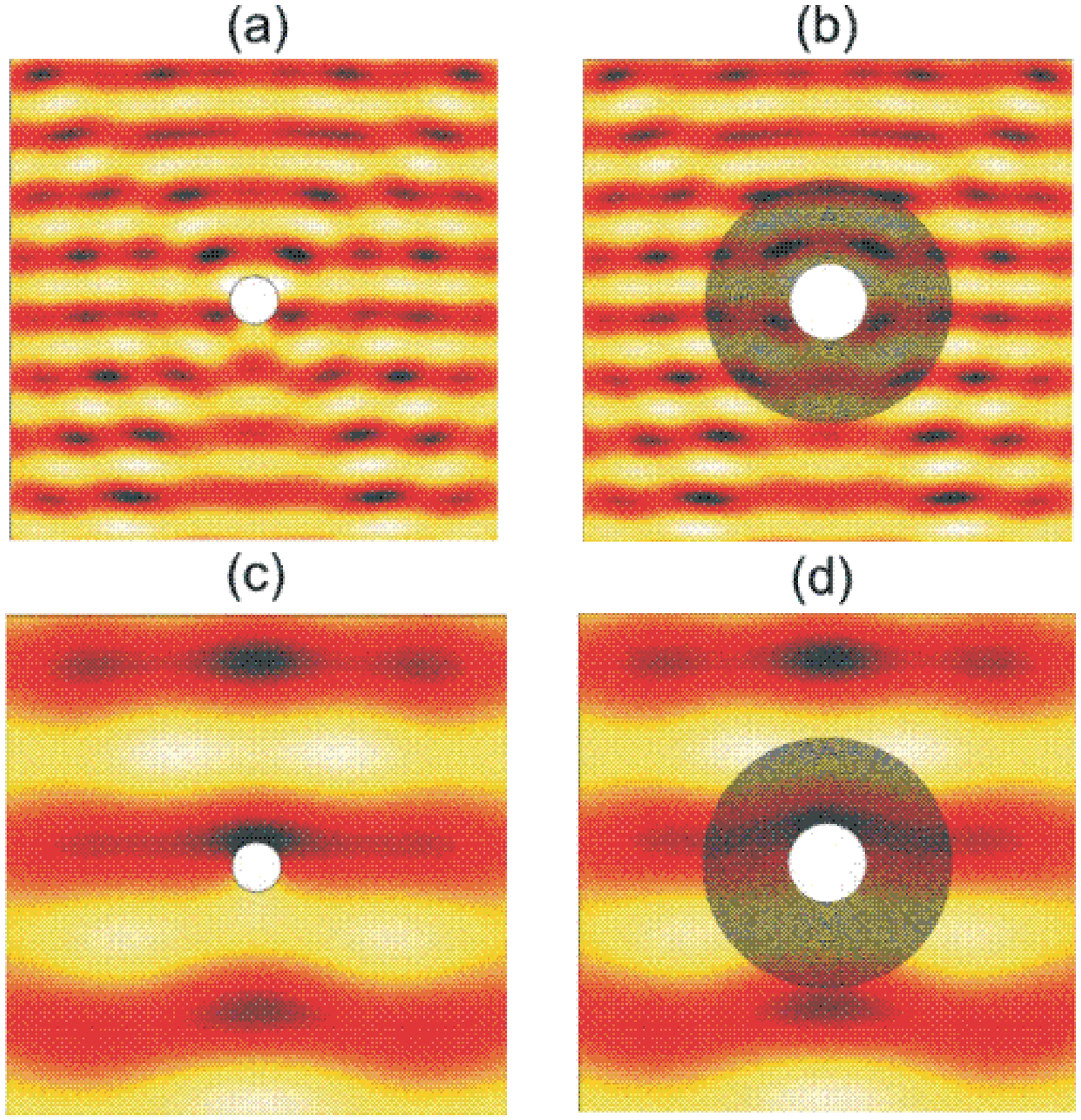}} %\includegraphics{cape2d1bis.eps}
\mbox{}\vspace{-0.4cm}\caption{(Color online) 2D plot of the real part
of the total pressure field $\Re{e}(p)$ for the same geometric and
acoustic parameters as in Fig. \ref{fig:multilayer}; Scattering by a
plane wave of wavelength $0.5$ (a-b) and $1.4286$ (c-d), incident
from above.} \label{fig:last}
\end{figure}

The acoustic parameters of the proposed layered circular carpet are
therefore characterized by a spatially varying scalar bulk modulus
$\rho$ and a spatially varying rank $2$ density tensor
$\underline{\underline{\rho}}$ given by (\ref{parameter}). We can
further simplify the problem by choosing reduced acoustic
parameters, so that the bulk modulus $\lambda$ is now constant, and
all the variation is reported on the density, see figure
\ref{fig:multilayer}. More precisely, $\rho_A$ varies in the range
$[0.1890 ; 0.5493]$ and $\rho_B$ varies in the range $[1.7987 ;
2.1472]$. We checked that this carpet is broadband as it works over
the range of wavelengths $\lambda\in[0.2,1.4286]$, see Fig.
\ref{fig:multilayer} and Fig. \ref{fig:last}: a multilayered carpet
of radius $1$ surrounding a rigid obstacle of radius $0.32$ scatters
waves just like a rigid obstacle of radius $R_0=0.2$. We note that
the lower bound for the range of working wavelengths corresponds to
the rigid obstacle we want to mimic.

\section{Conclusion}
In this paper, we have proposed some models of flying carpets which
levitate (or float) in mid-air (or mid-water). Such cloaks can be
built from acoustic metafluids: as explained by Pendry and Li in a
recent work, one can for instance emulate required anisotropic
density and heterogeneous bulk modulus with arrays of rigid plates
with a hemispherical sack of gaz attached to them \cite{pendrynjp}.
But other designs proposed by Torrent and Sanchez-Dehesa would work
equally well \cite{sanchez}. However, such flying carpets lead to
some approximate cloaking as they do not touch the ground (the inner
boundary of the carpet in the original design of Pendry and Li is
attached to the ground \cite{pendry-carpet}. Interestingly, other
authors also looked at quasi-cloaking with simplified carpets
\cite{kallos}. We have also explained how one can hide an object
located in the close neighborhood of a rigid circular cylinder,
which in some sense can be classified as an external cloaking
whereby a large scatterer hides smaller ones located nearby. Such an
ostrich effect (which buries its head in the sand) has already been
observed in the context of dipoles and even finite size obstacles
located closeby a cylindrical perfect lens which displays anomalous
resonances \cite{graeme,njpnico}. However, here the coating does not
contain any negatively refracting material, and this is an
anisotropy-led rather than plasmonic-type cloaking mechanism.
Actually, it is possible to use complementary media to cloak finite
size objects (rather than only dipoles) at a finite distance
\cite{chanprl}.

We have discussed some applications, with the sonar boats or radars
cases as typical examples. Another possible application would be
protecting parabolic antennas from the negative impact of their
`supporting cable'. The feasibility of such carpets is demonstrated
using a homogenization approach enabling us to design a
multi-layered acoustic metafluid leading to a mirage effect over a
finite range of wavelengths. We note that the route towards
anamorphism discussed in this paper is very different from the
proposal of Nicolet et al. \cite{polyjuice} whereby an anisotropic
heterogeneous object is placed within the coating of a singular
cloak to mimic the scattering of another object.
%Here, we do not need
%any extravagant material properties for the cloak to achieve
%mimetism.

Moreover, all computations hold for electromagnetic carpets built
with heterogeneous anisotropic permittivity and scalar permittivity,
which could be emulated using tapered waveguides as in \cite{igor}.
We therefore believe that the designs we proposed in this paper
might foster experimental efforts in approximate cloaking for both
acoustic and electromagnetic waves.
\section*{Acknowledgements}
AD and SG acknowledge funding from the Engineering and Physical
Sciences Research Council grant EPF/027125/1. 

%\bibliography{blibliopre}

%\section*{References}

\end{document}